\documentclass[preprint2]{aastex}
\pdfoutput=1
\usepackage{dcolumn}
\newcolumntype{d}{D{.}{.}{-1}}

\usepackage{graphicx,color,xspace}

\newcommand{\he}{Helium\xspace}                
                
\newcommand{\IBEX}{{\it IBEX}}
\newcommand{\Voyager}{{\it Voyager~1}}
\newcommand{\kms}{\ensuremath{\mathrm{km/s}}}
\newcommand{\nc}[1]{{ #1}}

\shorttitle{Return of the bowshock}
\shortauthors{Scherer and Fichtner}
\begin{document}

\title{The return of the bow shock}
\author{K. Scherer and H. Fichtner}
 \email{kls@tp4.rub.de,hf@tp4.rub.de}
 \affil{Institut f\"ur Theoretische Physik IV, Ruhr-Universit\"at Bochum, D-44780
   Bochum, Germany.}

\begin{abstract} 
  Recently it has been discussed whether a bow shock ahead of the 
  heliospheric stagnation region does exist or not. This discussion 
  was triggered by measurements indicating that the Alfv\'en speed and
  that of fast magnetosonic waves are higher than the flow speed of 
  the local interstellar medium (LISM) relative to the heliosphere and 
  resulted in the conclusion that there might exist either a bow wave or a 
  slow magnetosonic shock. We demonstrate here that including the He$^{+}$
  component of the LISM yields both an Alfv\'en and fast magnetosonic wave
  speed lower than the LISM flow speed. Consequently, the scenario of a 
  bow shock in front of the heliosphere as modelled in numerous simulations 
  of the interaction of the solar wind with the LISM remains valid.
\end{abstract}

\keywords{heliosphere, bow shock, local interstellar medium, helium abundance}
\maketitle

\section{\label{sec:intro}Introduction}
Recently, based on measurements made with the Interstellar Boundary
Explorer (\IBEX), \citet{McComas-etal-2012} concluded that the bow
shock in front of the heliosphere does not exist because the Alfv\'en
as well as the fast magnetosonic wave speeds are higher than the inflow
speed of the local interstellar medium (LISM) resulting in Mach
numbers $M$ of the order $0.9 < M < 1.0$.  While this was confirmed
with modelling by \citet{Zank-etal-2013}, \citet{Zieger-etal-2013}
showed that a so-called slow bow shock related to the slow
magnetosonic wave mode might exist, see also
\citet{Ben-Jaffel-etal-2013} for the fast shock.

These considerations did not take into account the presence of the
helium component of the LISM, however. While the significance of
helium for the large-scale structure of the heliosphere has been
revealed with simulations by \citet{Izmodenov-etal-2003} and
\citet{Malama-etal-2006}, corresponding multi-fluid modelling is not
yet standardly done. We demonstrate here that including the charged
helium component of the LISM is crucial for the comparison of the LISM
flow speed with the wave speeds and, thus, for the answer to the
question whether or not the interstellar flow is super-Alfv\'enic
and/or super-fast magnetosonic.


\section{The hydrogen and helium abundances in the LISM}

The neutral hydrogen and helium can be observed in-situ, either
directly \citep{Witte-2004,Bzowski-etal-2012, Moebius-etal-2012} or
indirectly via pickup ions in the solar wind
\citep{Bzowski-etal-2008, Gershman-etal-2013}. Because the heliopause
separates the solar from the interstellar plasma, the abundances of
protons and $He^{+}$ or other interstellar ions can only be determined
by remote measurements \citep[for an overview see][]{Jenkins-2013}
combined with modelling \citep{Slavin-Frisch-2008,Jenkins-2009}, see
also \citep{Barstow-etal-1997,Barstow-etal-2005}.  The largest
uncertainty in the observation concerns the magnetic field close to
the heliosphere, because such observations are, so far, only possible
on large galactic scales \citep[see, e.g.,][]{Frisch-etal-2012}.

In the following we use the values derived by
\citet{Slavin-Frisch-2008}, namely a proton number density of
$n_{p} = 0.07\pm0.005$\,cm$^{-3}$ and $n_{He^{+}} =
0.009\pm0.0027$\,cm$^{-3}$.
The latter value results from a neutral helium number density of
$n_{He} = 0.0151\pm0.003$\,cm$^{-3}$ combined with an ionization
fraction $X(He,He^{+})=0.4\pm0.1$.
The $He^{++}$ abundance in the
LISM  is negligible
\citep{Slavin-Frisch-2008}, as well as the ion abundance of other
elements. Thus, we take only the proton and $He^{+}$ ions into account
in what follows.

Note, that the sum of the number densities of the proton and helium charges
corresponds nicely to the recently observed electron number density
$n_{e}=0.08$\,cm$^{-3}$ observed with the plasma wave instrument
onboard \Voyager\ \citep{Gurnett-etal-2013}.

\section{\label{speeds}Characteristic speeds in a multi-ion plasma}

In order to quantify the effect of the charged helium component in the
LISM, we compute the sound, Alfv\'en and fast magnetosonic wave speeds
for both a pure proton-electron plasma and a proton-He$^+$-electron
plasma. For the respective sounds speeds one has
\citep[see,
e.g.,][]{Fahr-etal-1997,Fahr-Rucinski-1999,Izmodenov-etal-2003}:
\begin{eqnarray}
  \label{eq:1}
  v_{s_p}^{2} &=& \frac{\gamma kT}{m_{p}} \\ 
  v_{s}^{2} &=&\frac{\gamma\sum\limits_{i}
    P_{i}}{\sum\limits_{i}\rho_{i}} =
  \frac{\gamma(P_{p}+P_{He^{+}})}{n_{p}m_{p}+n_{He^{+}}m_{He}} \nonumber\\
  &\approx&
  \frac{n_{p}+n_{He^{+}}}{n_{p}+4n_{He^{+}}} \frac{\gamma kT}{m_{p}}  =
  \frac{n_{p}+n_{He^{+}}}{n_{p}+4n_{He^{+}}} v_{s_p}^{2}\nonumber\\ 
 &=&\frac{1+\mu}{1+4\mu}v_{s_p}^{2}
\end{eqnarray}
where $P_{i}$, $\rho_{i}$ and $m_{i}$ denote the pressure, mass
density, and mass of protons ($i=p$) and helium ions ($i=He^+$),
respectively, $\gamma=5/3$ and $k$ is the Boltzmann constant. For the
last equality the ratios $m_{He^+}/m_p\approx 4$ and
$\mu = n_{He^+}/n_p$ have been used.

Similarly, the respective Alfv\'en speeds read
\citep[e.g.,][]{Marsch-Verscharen-2011}:
\begin{eqnarray}
  \label{eq:2}
  v_{A_{p}} &=&  \frac{B}{\sqrt{4\pi n_{p}m_{p}}} \\
  v_{A} &=& \frac{B}{\sqrt{4\pi \sum\limits_i \rho_{i}}}\nonumber\\
  &=&\frac{B}{\sqrt{4\pi (n_{p}m_{p}+n_{He^{+}}m_{He} )}} \nonumber\\ 
  &\approx&
  \frac{B}{\sqrt{4\pi m_{p}}}\frac{1}{\sqrt{n_{p}+4n_{He^{+}}}}\nonumber\\ 
  &=& v_{A_{p}}\frac{1}{\sqrt{1+4\mu}}
\end{eqnarray}
with $B$ being the strength of the magnetic field.

From formula (1) to (4) the fast (fw, $+$) and slow (sw, $-$)
magnetosonic wave speeds can be obtained in the form
\citep[e.g.,][]{Boyd-Sanderson-2003}:
\begin{eqnarray}
  \label{eq:3}
  v_{fw_p,sw_p}^{2}&=&\frac{1}{2}\left[v_{A_{p}}^{2}+v_{s_{p}}^{2} \pm \right.\nonumber\\
                   & & \!\!\!\!\!\!\!\!\!\! \left. \sqrt{\left(v_{A_{p}}^{2}+v_{s_{p}}^{2}\right)^{2 } - 4
      v_{A_{p}}^{2}v_{s_{p}}^{2} \cos\vartheta} \right] \\
  v_{fw,sw}^{2}&=& \frac{1}{2}\left[v_{A}^{2}+v_{s}^{2} \pm \right.\nonumber\\
    && \!\!\!\!\!\!\!\!\!\! \left.\sqrt{\left(v_{A}^{2}+v_{s}^{2}\right)^{2 } - 4
      v_{A}^{2}v_{s}^{2} \cos\vartheta} \right] 
\end{eqnarray}
where $\vartheta$ denotes the angle between the propagation direction
of a magnetosonic wave and the magnetic field. Since we are only
interested in those waves traveling the shortest distance to the
heliosphere, i.e.\ those traveling in the direction of the inflow
velocity, $\vartheta$ is taken as the angle between the inflow
velocity and the magnetic field direction.

\section{\label{lismspeeds}Characteristic speeds in the LISM}

The LISM can be characterized with a temperature (for both protons and
helium ions,
\nc{see, however, the discussion at the end of section~6}) of
$T =6300\pm340$\,K, a speed of $v_{LISM} = 23.2$~km/s, and a magnetic
field strength of about 3\,$\mu$G
\citep{Frisch-etal-2012,McComas-etal-2012b}.  These `most likely
values' correspond to an equipartition between the magnetic field
pressure and the total pressure, i.e.\
$B^{2}/(8\pi)\approx (n_{p}+n_{He^{+}})kT\approx
3.55\cdot10^{-13}$\,erg/cm$^{3}$.
Furthermore, we assume that the heliosphere is a stationary structure
with respect to the LISM and, thus, only the given interstellar
parameters are needed to check on the existence of the bow shock.  The
situation becomes more complicated when taking into account dynamic
variations due to the solar cycle, which can affect the position of
the bow shock \citep[e.g.,][]{Scherer-Fahr-2003b}.

\begin{figure}[t!]
  \centering
  \includegraphics[width=1.0\columnwidth]{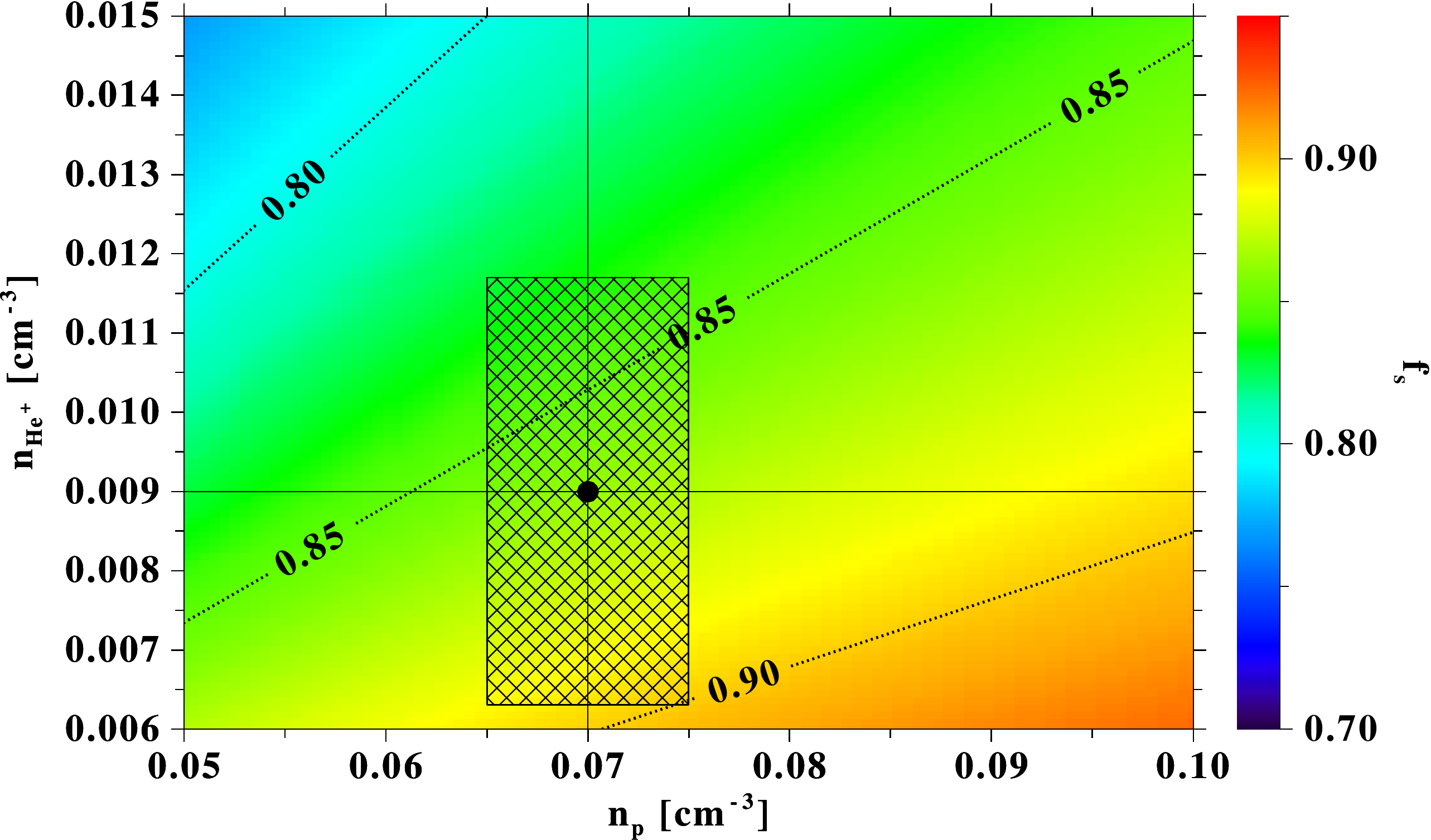}
  \caption{The multi-ion sound speed normalized to the proton sound
    speed as a function of the number densities $n_p$ and
    $n_{He^{+}}$. The black lines denote the `most likely values' for the
    proton and $He^{+}$ number densities. \nc{The dotted lines are the
    contours for $f_{s}=0.8,0.85,0.9$}}
  \label{fig:1}
\end{figure}

\begin{figure}[t!]
  \centering
  \includegraphics[width=1.0\columnwidth]{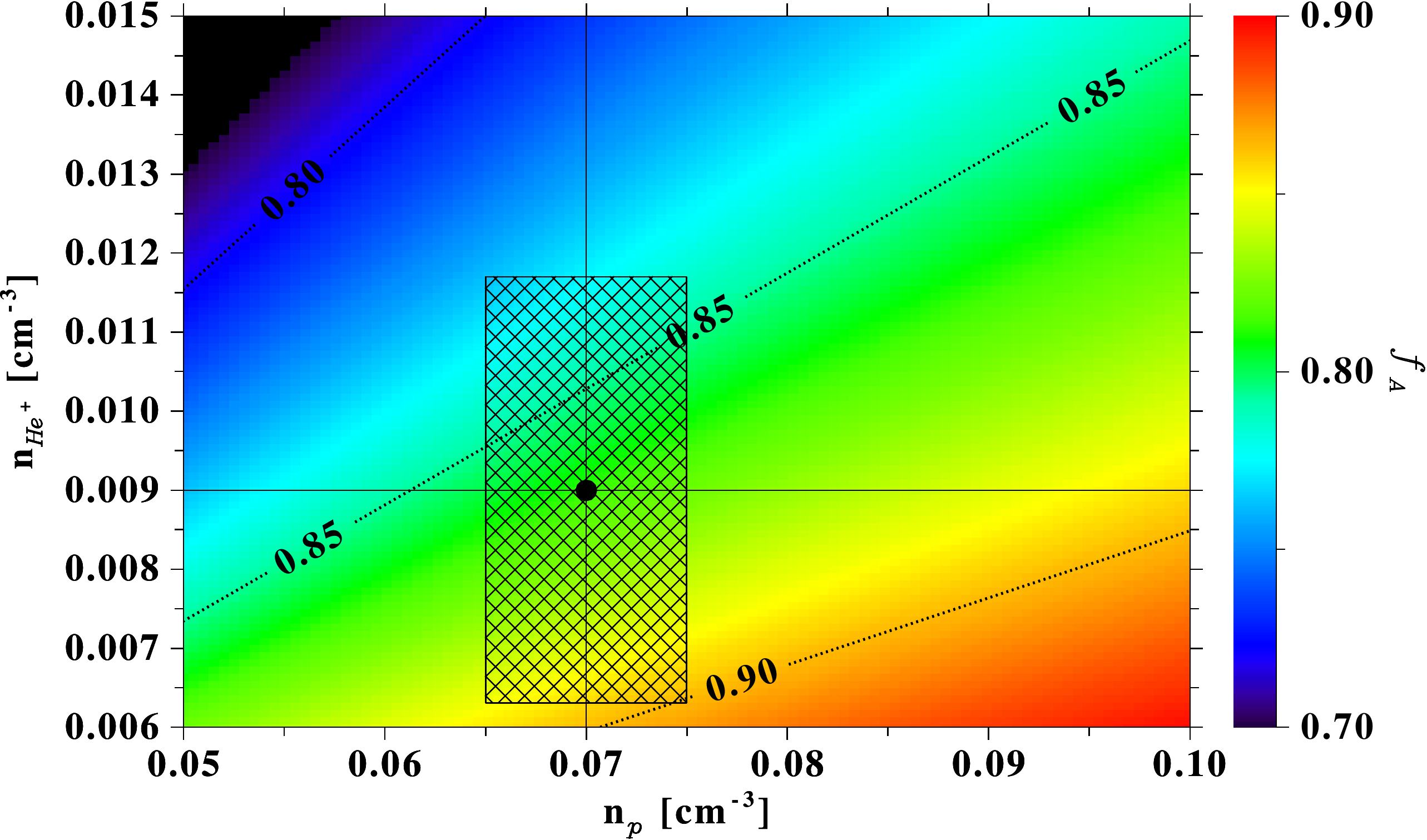}
  \caption{Same as Fig.~\ref{fig:1}, but for the Alfv\'en speeds.}
  \label{fig:2}
\end{figure}

In principle one has a five dimensional parameter space
$(n_{p},n_{He^{+}},B,T,\vartheta)$. Here we will concentrate on the
dependence of the speeds on $n_{He^+}$ and we will discuss the
significance of the uncertainties in the other quantities in the next
section.

In Fig.~\ref{fig:1} the ratio $f_{s}=v_{s}/v_{s_p}$ of the
multi-ion sound speed to the proton sound speed  \nc{is plotted}  as a function of the
number densities $n_p$ and $n_{He^+}$. Fig.~\ref{fig:2} shows the
correspding plot for the Alfv\'en speeds, i.e.\ $f_{A}=v_{A}/v_{A_p}$. The
black lines indicate those speeds derived from the above most likely values,
which are given together with magneto-sonic speeds in
Table~\ref{tab:1}.

\begin{table*}[t!]
  \centering
 \setlength{\tabcolsep}{1mm}
\setlength{columnseprule}{1mm}
  \begin{tabular}{l|ddd|ddd|d}
  & \multicolumn{3}{c|}{protnons-only} &
\multicolumn{3}{c|}{proton+helium} & \\ 
$i$     
& \multicolumn{1}{c}{\ensuremath{v_{i,p}}} 
& \multicolumn{1}{c}{\ensuremath{\pm\Delta v_{i,p}}} 
& \multicolumn{1}{c|}{\ensuremath{\Delta v_{i,p}/v_{i,p}}} 
& \multicolumn{1}{c}{\ensuremath{v_{i}}} 
& \multicolumn{1}{c}{\ensuremath{\pm\Delta v_{i}}}
& \multicolumn{1}{c|}{\ensuremath{\Delta v_{i}/v_{i}}}  
& \multicolumn{1}{c}{\ensuremath{f_{i}=v_{i}/v_{p,i}}} \\
& \multicolumn{1}{c}{km/s} & \multicolumn{1}{c}{km/s} & \multicolumn{1}{c|}{\%} & \multicolumn{1}{c}{km/s} & \multicolumn{1}{c}{km/s} & \multicolumn{1}{c|}{\%} & \\
\hline
$s$          &  9.35 &  0.22 & 0.022 &  8.07 &  0.64  & 0.079  & 0.86 \\
$A$          & 24.73 &  4.20 & 0.17  & 20.10 &  4.02  & 0.20   & 0.81 \\
\hline
$fw(0^{o})$  & 24.73 &  8.41 & 0.34  & 22.85 &  7.24  & 0.36  & 0.81 \\
$fw(45^{o})$ & 25.30 &  8.37 & 0.33  & 20.62 &  7.12  & 0.35  & 0.82 \\
$fw(90^{o})$ & 26.44 &  9.44 & 0.36  & 21.66 &  7.27 &  0.34  & 0.82 \\
$sw(0^{o})$  &  9.35 &  0.033 & 3.7\ensuremath{\cdot 10^{-4}} &  8.07 &  0.004 & 5.5\ensuremath{\cdot 10^{-4}} & 0.86 \\
$sw(45^{o})$ &  7.68 & 9.409 & 1.22   &  6.61 &  0.004 & 6.8\ensuremath{\cdot 10^{-4}} & 0.86 \\
$sw(90^{o})$ &  0.00 & \ensuremath{0.003^{*}} & &  0.00 & \ensuremath{0.005^{*}}      \\
  \end{tabular}
  \caption{The most likely speeds and its errors for a proton-only and a
    proton-helium fluid. The last column gives the ratios between the
    proton-only to the proton-helium speeds
    $f_{i}=v_{i,p}/v_{i}$.}
  \label{tab:1}
\end{table*}

\begin{figure}[t!]
  \centering
  \includegraphics[width=1.\columnwidth]{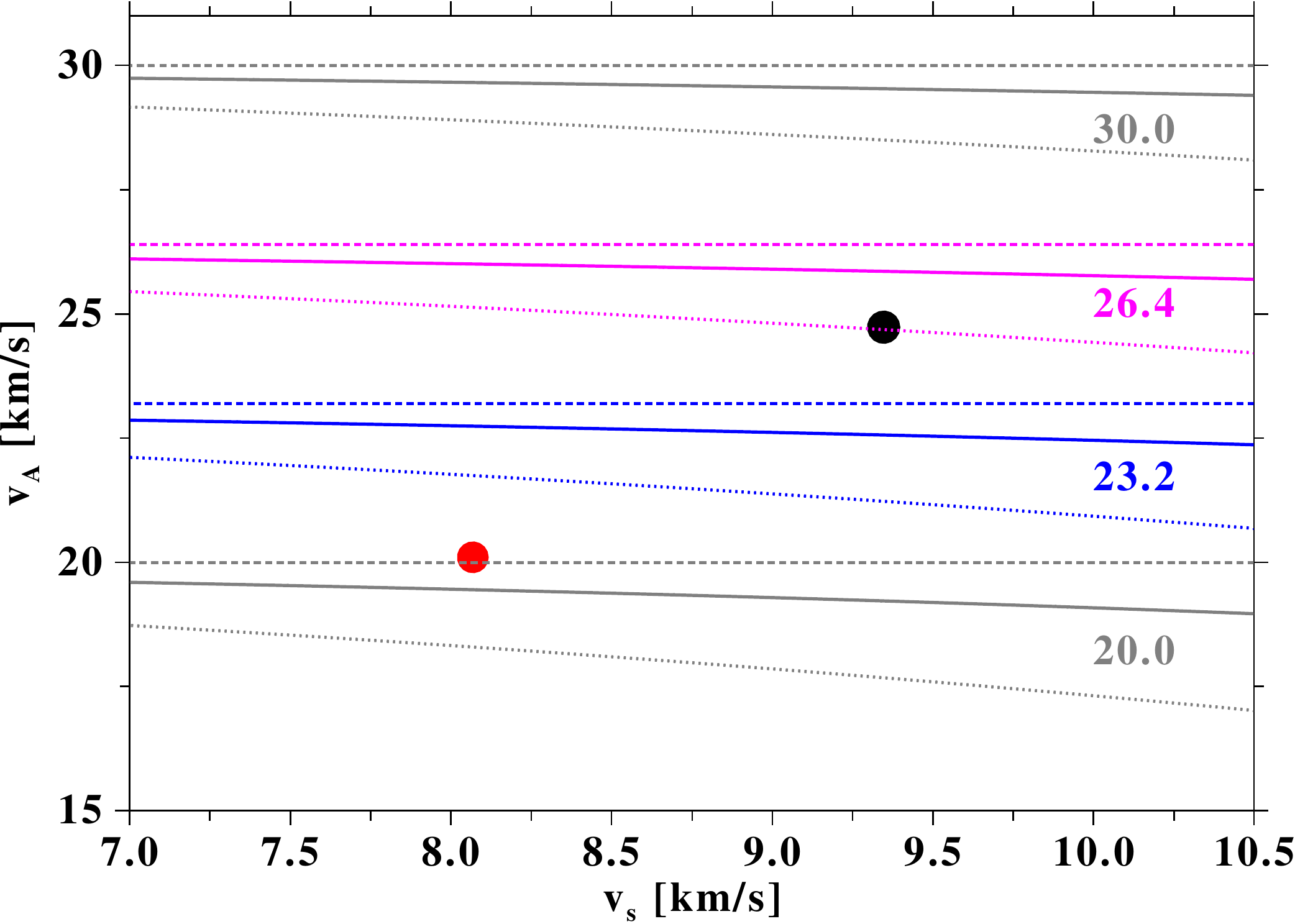}
  \caption{\nc{Fast magnetosonic speed (FMS) as function of
      $v_{s},v_{A}$ and $\vartheta$. The line types are the contours
      for the angles $\vartheta=0^{o}$ (dashed), $\vartheta=45^{o}$
      (solid), and $\vartheta=90^{o}$ (dotted). The numbers give the
      corresponding FMSs.  The black dot denotes the FMS for the `most
      likely values' $v_{s_{p}},v_{A_{p}}$ for the protons-only, and
      the red dot is that for $v_{s},v_{A}$, i.e.\ including the
      $He^{+}$ contribution.  The blue lines are the contours for a
      FMS of 23.2\,\kms, while the magenta lines that for 26.4\,\kms,
      which are identical with the respective LISM speeds as observed
      by \IBEX\ \citep{McComas-etal-2012} and {\it Ulysses}
      \citep{Witte-2004}. In the latter case the most likely value for
      the proton-only fluid lies below the solid line and thus the
      fluid is super-magnetosonic, while for the speed of 23.2\kms\
      the lines always are below the black dot and thus the fluid is
      sub-magnetosonic. However, for the multi-fluid case, both LISM
      speeds are above the red dot, and hence the flow is super-fast
      magnetosonic.  The gray lines give the values for FMS of
      20\,\kms\ and 30\,\kms, respectively.  }}
  \label{fig:3}
\end{figure}

In Fig.~\ref{fig:3} the fast magnetosonic wave speeds for the angles
$\vartheta \in\{0^{o},45^{o},90^{o}\}$ are plotted. The red dot
represent the values for the proton\nc{-only} sound and Alfv\'en speeds given
above, while the white dot determines the magnetosonic wave speed for
the multi-ion speeds.

From Fig.~\ref{fig:3} we can deduce that for the most likely set of interstellar
parameters the fast magnetosonic speed as well as the Alfv\'en and
sound speeds are below the 23.2~km/s line and, thus, a bow shock must
be expected to exist. In the following section we discuss in what
range the error bars are.

\section{\label{eroor}Uncertainties of the characteristic speeds}
The error $\Delta$ for a function $g(x_{1},...,x_{n})$ is given by:
\begin{eqnarray}
  \label{eq:error}
  \Delta g = \sqrt{\sum\limits_{i=1}^{n} \left(\Delta x_{i}
    \frac{\partial g(x_{1},...,x_{n})}{\partial x_{i}}\right)^{2} }
\end{eqnarray}
\nc{The relative errors $\Delta g /g$ for the}  sound speeds
$v_{s_{p}}$ and $v_{s}$ yields:
\begin{eqnarray}
  \label{eq:vs_ero}
\frac{\Delta v_{s_{p}}}{v_{s_{p}}} &=& \frac{\Delta T}{2 T} \\\nonumber
  \frac{\Delta v_{s}}{v_{s}} &=& \sqrt{\left(\frac{\Delta T}{2 T}\right)^{2}+
    \frac{9}{4\mu^{2}}
    \frac{\left(\frac{\Delta  n_{p} }{n_{p}}     \right)^{2}+
          \left(\frac{\Delta  n_{He}}{n_{He^{+}}}\right)^{2}}
{(1+\mu)^{2}(1+4\mu)^{2}}
} 
\end{eqnarray}
and, analogously, for the Alfv\'en speeds $v_{A_{p}}$ and $v_{A}$:
\begin{eqnarray}
  \label{eq:va_ero}
 \frac{\Delta v_{A_{p}}}{v_{A_{p}}} &=& 
  \sqrt{\left(\frac{\Delta B}{B}\right)^{2}+\left(\frac{\Delta n_{p}}{2n_{p}}\right)^{2}}\\\nonumber
  \frac{\Delta v_{A}}{v_{A}} &=& 
  \sqrt{\left(\frac{\Delta B}{B}\right)^{2} 
    +\frac{\left(\frac{\Delta n_{p}} {n_{p}}\right)^{2} +16 
           \left(\frac{\Delta n_{He}}{n_{He^{+}}}\right)^{2}\mu^{2}}{4(1+4\mu)^{2}}
} 
\end{eqnarray}
The corresponding expressions for the magnetosonic speeds
$v_{fw_p,sw_p}$ and $v_{fw,sw}$ are very clumsy and were, therefore,
calculated with the help of computer algebra system
`wxmaxima'\footnote{(http://sourceforge.net/projects/wxmaxima/)} and
are not given here.

With the `most likely values' as given above, i.e.\
$\Delta T/T = 340/6300\approx 0.054$,
$\Delta n_{p}/n_{p}=0.005/0.07\approx 0.07$,
$\Delta n_{He^{+}}/n_{He^{+}}=0.009/0.015=0.02$, and
$\Delta B /B=0.5/3=0.17$.
\nc{we can calculate the relative uncertainties in the speeds
  $\Delta v_{i}/v_{i}$ ($i\in\{s,A,fw,fs\}$) for the sound,
  Alfv\'en, fast and slow magnetosonic speeds, given in
  Table~\ref{tab:1} together with the extreme values
  $\vartheta=0^{o},90^{o}$.}
The reason for the large errors in the Alfv\'en speed is the
uncertainty in the strength of the interstellar magnetic field,
\nc{which determines those of the magnetosonic speeds.} 
Obvioulsy, these uncertanities do not influence the main conclusion
for the most likely values: The interstellar flow speed is likely to
be super-fast magnetosonic and, thus, a fast bow shock must be
expected to exist.

To complete the assessment, we would like to remark that if the
`traditional' value of 26.3\,km/s for LISM inflow speed were correct,
only magnetic field values above $>3.5\,\mu$G would remove the bow
shock. Such rather extreme values of the magnetic field are discussed
in \citet{Zieger-etal-2013} but are not favored by other authors
\citep[see, e.g.,][]{Zank-etal-2013}.

\section{Compression ratio of the bow shock}
Given that it is likely that a fast bow shock exists it is interesting
to estimate its strength.  In ideal MHD the compression ratio $s$ at a
shock fulfils the equation \citep{Kabin-2001}:
\begin{eqnarray}
  \label{eq:s1}
0= (A_{1}^{2}-s)^{2} \left[A_{1}^{2}-\frac{2 s S_{1}^{2}}{(s+1-\gamma (s-1))}\right]\\\nonumber  
-s k^{2}_{1} A_{1}^{2} \left[\frac{(2 s-\gamma (s-1))}{(s+1-\gamma(s-1))}A_{1}^{2}-s\right]
\end{eqnarray}
with the abbreviations
$ A_{1} = {M_{A_{1}} \cos\alpha}/{\cos(\alpha-\vartheta)}$,
$ S_{1} = {M_{A_{1}}}/{M_{s_{1}}\cos(\alpha-\vartheta)}$, and
$ k_{1} = \tan{(\alpha-\vartheta)}$.  Here $\alpha$ is the angle
between the inflow direction and the shock normal, $\vartheta$ is the
angle between the inflow and the magnetic field direction, and
$M_{A_{1}}, M_{s_{1}}$ are the Alfv\'enic and the sound wave Mach
number upstream of the shock.
\begin{figure}[t!]
  \centering
  \includegraphics[width=1.0\columnwidth]{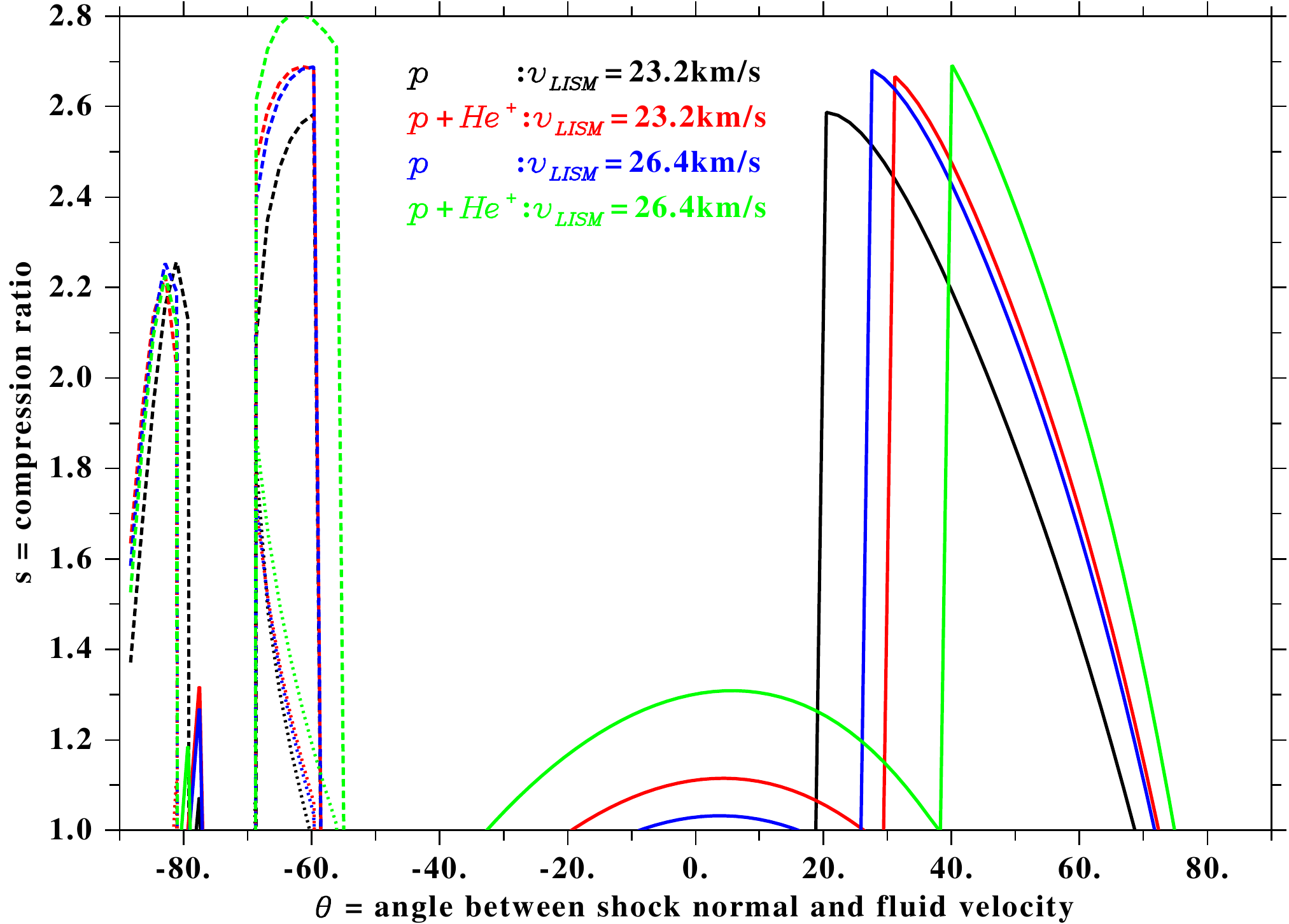}
  \caption{The compression ratio along the bow shock. The dashed and
    dotted lines denote the so-called intermediate shocks.}
  \label{fig:6}
\end{figure}
While a detailed analysis of the MHD shock structure -- that should
even be influenced by the presence of neutrals coupled via
charge-exchange to the plasma \citep[see, e.g.,][]{Lu-etal-2013} --
would go far beyond the scope of this article, we nonetheless show with
Fig.~\ref{fig:6} that even an ideal MHD bow shock ahead of the
heliopause has a complicated structure.

The figure gives a graphical representation of the solutions of
equation~\ref{eq:s1}, which can be reformulated as a cubic polynomial
in the compression ratio $s$. The three groups of curves correspond
to the three solutions and can be interpreted \nc{from left to right
  as intermediate,  classical as well as
slow shocks} \citep{Kabin-2001}.  These solutions suggest that
especially towards the flanks of the heliosphere the bow shock can be
characterized as an intermediate rather than a classical one.

From the figure one expects that, most likely, a shock transition
parallel to the inflow direction ($\alpha = 0^{o}$) exists, at least,
when the $He^{+}$ component in the interstellar medium is taken into
account \citep{Scherer-etal-2013}.

\nc{Because an MHD shock structure can only be determined ``a
  posteriori'', a detailed analysis of the MHD-shock behavior requires
  an MHD model including \he. Such is not available at the moment,
  except that described by \citet{Izmodenov-etal-2003} and
  \citet{Malama-etal-2006} which are HD-models. Moreover, as was shown
  by \citet{Zank-etal-2013} energetic neutrals (ENAs) generated in the
  un-shocked solar wind can leak into the LISM and heat the latter.
  Because the charge exchange process does not change the number
  density of protons, the Alfv\'en speed $v_{A}$ is not affected,
  neither is the denominator in the sound speed given in Eq. 1. Only
  the nominator of the latter changes, and thus the sound speed will
  increase slightly (when we assume that the number density of ENAs
  and, hence, that of newly born ions is small compared to the
  interstellar proton density). As one can read off from Fig. 3 the
  dependence of the fast magnetosonic wave is weak as long as
  $v_{s}<v_{A}$. Hence, the overall conclusion remains: the bow shock
  is likely to exist.

}

\section{Conclusion} 

We have demonstrated that the Alfv\'en and the fast magnetosonic wave
speeds are -- despite the uncertainties in the values characterizing
the local interstellar medium -- lower than the inflow speed of the
interstellar medium and, thus, that a fast bow shock most likely exists. We
arrived at this conclusion by explicitly taking into the account the
effect of interstellar helium on the characteristic wave speeds. This
result re-emphasises the need of including $He^{+}$ ions in the
modeling of large-scale heliospheric structure.

We have also illustrated that the structure of the bow shock is more
complicated than than that of a purely hydrodynamic one. In any case,
the existence of the bow shock depends strongly on the strength of the
local interstellar magnetic field, which will hopefully be measured in
the near future by the \Voyager\ spacecraft that has recently crossed
the heliopause \citep{Gurnett-etal-2013}.

\begin{acknowledgements}
  KS and HF are grateful to the
  \emph{Deut\-sche For\-schungs\-ge\-mein\-schaft, DFG\/}, for funding
  the projects FI706/15-1 and SCHE334/10-1. 
\end{acknowledgements}


\end{document}